# Absence of Orbital Hall Magnetoresistance in Nonmagnet/Ferromagnet Bilayers with Large Orbital Torque

Yumin Yang[1,2], Wenqi Xu[1,2], Na Lei[3,4*], Zhicheng Xie[5], Dahai Wei[1,2*] and Jianhua Zhao[4]

[1]*State Key Laboratory of Semiconductor Physics and Chip Technologies, Institute of Semiconductors, Chinese Academy of Sciences, Beijing 100083, China*

[2]*College of Materials Science and Opto-Electronic Technology, University of Chinese Academy of Sciences, Beijing 100049, China*

[3]*Fert Beijing Institute, MIIT Key Laboratory of Spintronics, School of Integrated Circuit Science and Engineering, Beihang University, Beijing, 100191, China*

[4]*National Key Laboratory of Spintronics Hangzhou International Innovation Institute Beihang University, Hangzhou 311115, China*

[5]*Southwest Institute of Technical Physics, Chengdu, 610041, China*

E-mail: na.lei@buaa.edu.cn; dhwei@semi.ac.cn

## ABSTRACT

We report the absence of orbital Hall magnetoresistance (OMR) in nonmagnet/ferromagnet bilayers, challenging the general assumption that orbital transport mimics spin transport. Despite the observation of giant orbital torques, confirming the generation of orbital currents, thickness-dependent magnetoresistance measurements reveal that the signal is dominated by the intrinsic magnetoresistance of the ferromagnet and current shunting, with no discernible OMR contribution. We attribute this contradiction to the distinct transport properties of orbital compared with spin. Orbital currents undergo isotropic bulk absorption in the ferromagnet rather than anisotropic interfacial reflection required for OMR. Furthermore, we find that texture-induced magnetoresistance and self-torques in Ni-based bilayers can generate misleading signals, suggesting that caution is required when employing Ni in orbitronic studies. These findings clarify the distinct physical rules governing orbital transport and provide a simple method to distinguish spin and orbital currents.

# Main Text

Over the past decade, spin-related phenomena in systems with strong spin-orbit coupling (SOC), particularly the spin Hall effect (SHE) [1–3] and spin-orbit torque (SOT) [4,5], have attracted considerable attention. Recently, research on nonequilibrium orbital angular-momentum (OAM) transport, namely orbital current, has expanded the scope of spintronics by harnessing the orbital degree of freedom [6,7]. Distinct from spin, OAM exhibits stronger lattice coupling [8] but weaker exchange interaction with the magnetic moment [9,10], offering a distinct channel for angular momentum flow. Moreover, efficient magnetic manipulation via the orbital Hall effect (OHE) [11–14] and orbital torque [15–17] has been demonstrated even in light elements with weak SOC, greatly expanding the material landscape of spin-orbitronics beyond heavy metals. However, many interpretations of orbital-related phenomena implicitly assume that orbital transport simply follows the same phenomenological rules as spin transport, that is, the "orbital version" of spin-related effects [6]. Given the fundamentally different nature of OAM [13,18], this spin-orbital analogy might hinder the rational design of efficient spin-orbitronic devices and lead to the misidentification of spin-related signatures in light metals as orbital effects.

A typical distinction between spin and orbital transport lies in their propagation characteristics. Orbital currents exhibit significantly longer transport ranges within ferromagnets [19–22], potentially enabling non-local control of magnetization. Beyond bulk transport, the interfacial transport at the nonmagnet/ferromagnet (NM/FM) interface may reveal deeper discrepancies. In traditional NM/FM bilayers with strong SOC, a typical phenomenon is spin Hall magnetoresistance (SMR) [23–25]. Similarly, orbital Hall magnetoresistance (OMR), the orbital counterpart to SMR, is proposed to interpret the SMR-like magnetoresistance (MR) behaviors in NM/FM bilayers with weak SOC [26,27]. However, current interpretations of the results remain controversial [28,29]. Moreover, orbital transport parameters inferred from OMR [26] differ significantly from those established via orbital torque [19]. These results indicate that orbital transport at the NM/FM interface cannot be simply likened to spin transport.

Conventionally, for SMR, spin current generated by NM undergoes reflection at NM/FM interface, which depends on the relative orientation of spin polarization $\sigma_{\text{spin}}$ and magnetization $M$. The different spin reflection between $\sigma_{\text{spin}} \parallel M$ and $\sigma_{\text{spin}} \perp M$ subsequently generates various charge currents via inverse SHE, thereby inducing anisotropic resistance [23,24] (see Fig. 1a-b). A simple analogy suggests that strong OHE or orbital torque in NM/FM bilayer should naturally be accompanied by a sizable OMR. However, considering that orbital current couples weakly to the magnetic moment [9,10] and has longer transport range in FM [19–22], we propose an alternative physical picture. When orbital polarization $\sigma_{\text{orb}} \perp M$, the FM absorbs the orbital current and generates orbital torque, as shown in Fig. 1c; whereas $\sigma_{\text{orb}} \parallel M$, the orbital current dissipates within the FM instead of being strongly reflected, as illustrated in Fig. 1d due to the weak coupling. If the OAM undergoes isotropic bulk absorption rather than interface reflection, the OMR required anisotropic reflection would vanish. Consequently, the OMR, involving multiple transport processes, provides an ideal testbed for characterizing interfacial orbital transport.

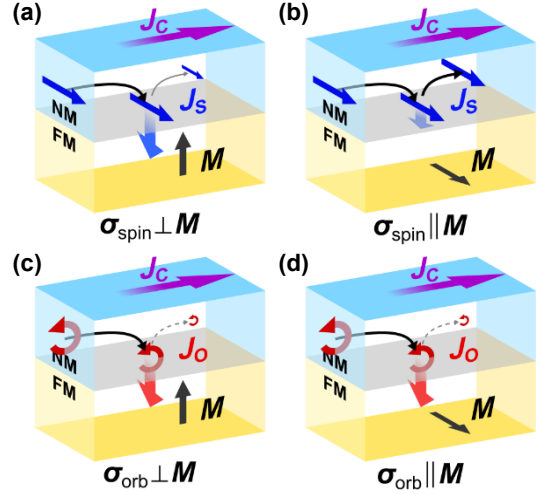

Fig. 1. (a-d) Schematic comparison of spin (top) and orbital (bottom) transport at the NM/FM interface. Where, blue straight and red curved arrows represent spin and orbital angular momentums respectively, black arrows denote the magnetic moment $M$, the purple arrows indicate charge current $J_C$. (a, c) For polarization $\sigma \perp M$, both spin current $J_S$ and orbital current $J_O$ transfer angular momentum to FM. (b) For $\sigma \parallel M$, spin currents are reflected, generating SMR. (d) In contrast, orbital currents are strongly absorbed by FM, suppressing the reflection required for OMR.

In this Letter, we demonstrate that the magnetization-independent orbital current reflection and absorption lead to the absence of OMR through a systematic study of NM/FM bilayers, where we use Ru and Ti with weak SOC as orbital current sources and Ni, Co, and Py as FMs. Considering the complexity and multiplicity of MR mechanisms, we investigate the change of MR ratio with respect to NM thickness $t_{NM}$, and carefully control the FM thickness $t_{FM}$ and NM/FM interface. However, after rigorously excluding FM's intrinsic mechanisms and NM's shunting effects, the OMR signal is vanishingly small across all samples. Crucially, the observation of giant orbital torques confirms the existence of orbital current in the same devices. This supports our hypothesis that orbital transport is dominated by isotropic bulk absorption in FM, in contrast to the anisotropic interfacial reflection mechanism of spins. Moreover, we point out the complexity and shortcomings of Ni-based heterojunctions in the study of orbital-related effects. Our findings not only clarify the distinction between orbital and spin, but also provide a simple method to distinguish them.

In the experiment, all samples were prepared by magnetron sputtering with $10^{-9}$ Torr base pressure at room temperature. The heterojunction films were deposited on sapphire substrates with $SiO_2(10)Ti(2)$ buffer layers and $Ti(1)SiO_2(10)$ capping layers; the numbers in parentheses represent the thickness in nanometers, and the layer sequence corresponds to the deposition order. To minimize the variations among samples, shutters are employed to produce samples of varying thicknesses in a single growth process. Then, the films were fabricated into devices using standard photolithography, ion-beam etching, and lift-off.

In order to evaluate the OMR, angular-dependent magnetoresistance (ADMR) measurements were systematically performed on all samples. A charge current was applied along the x-axis of the Hall bar devices, while an external magnetic field $H$ was rotated within the yz-plane as illustrated in Fig. 2a. To minimize the angular deviation between magnetization direction $M$ and $H$, magnetic fields of 5T and 9T were applied for Ni-based and Co-based samples, respectively. The normalized

ADMR $(R(\beta) - R_0)/R_0$ for Ru(t)/Co(5) series are shown in Fig. 2b, and can be described by $R(\beta) = R_0 + \Delta R \cos^2 \beta$, which is similar to that for SMR. However, the appearance of such SMR-like ADMR does not constitute evidence for OMR. It can arise from interfacial SOC [29–31], geometric-size effect and film texture [32–34], all of which are sensitive to $t_{FM}$. Thus, one possible approach to prove the existence of OMR is the $t_{NM}$-dependence of the MR ratio $\Delta R/R_0$. Fig. 2c presents the $\Delta R/R_0$ for Ru(t)/Co(5) series, which monotonically decreases from $8\times10^{-4}$ to $5\times10^{-5}$ with $t_{Ru}$ increasing, where $\Delta R = R_\perp - R_\parallel$, and $R_\perp$ and $R_\parallel$ represent the $R(\beta)$ with the magnetic field perpendicular ($\beta$=0°) and parallel ($\beta$=90°) to the film, respectively. However, the monotonical decay of $\Delta R/R_0$ is likely dominated by the MR mechanisms within FM layer [34] together with the shunting effect of NM layer, rather than exhibiting SMR-like behavior with an asymmetric single-peak (increases rapidly with $t_{NM}$ and then gradually decreases as the predicted OMR plotted in Fig. 2c, to be discussed later). According to the shunting model, the variation of $\Delta R/R_0$ with $t_{NM}$ caused by current shunting can be approximated to first order as,

$$\frac{\Delta R}{R} = r_0 / \left(1 + \frac{t_{NM}\rho_{FM}}{t_{FM}\rho_{NM}}\right) \quad (1)$$

where $\rho_{NM}$ and $\rho_{FM}$ are the resistivities of NM and FM, and $r_0$ is the MR ratio at $t_{NM}$=0. Here, only $r_0$ was treated as a free parameter and the resistivities were obtained from $R_0$, yielding $\rho_{Ru}$=22.4 μΩ·cm and $\rho_{Co}$=167.3 μΩ·cm. From Fig. 2c, the shunting model reproduces well the experimental dependence of $\Delta R/R_0$ on $t_{Ru}$ and significantly differs from predicted OMR. This result indicates that the MR is dominated by the mechanisms of FM single layer rather than OMR.

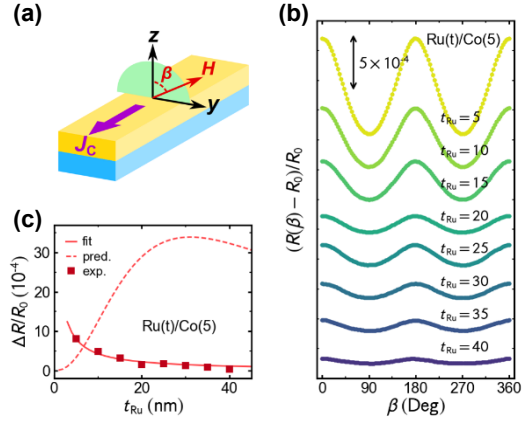

Fig. 2. (a) Schematic of the ADMR measurement with the external magnetic field $H$ rotating in the yz-plane ($\beta$ angle). (b) Normalized angular-dependent resistance $(R(\beta) - R_0)/R_0$ for Ru(t)/Co(5) bilayers. (c) MR ratio $\Delta R/R_0$ as a function of $t_{Ru}$. The solid line represents the fit to the shunting model, while the dashed line indicates the predicted OMR signal using SMR-like model.

The absence of the OMR might also be attributed to the weak orbital current in the NM/FM bilayer. To rule out this possibility, we first conducted spin-torque ferromagnetic resonance (ST-FMR) experiments to measure the orbital torque, thereby verifying the presence of sizable orbital current. A modulated radio frequency (RF) current is applied into ST-FMR devices, a swept in-plane magnetic field $H$ is applied at various azimuthal angles $\varphi$ to probe the angular dependence of the signal, and the rectified voltage signal $V_{DC}$ is detected using a lock-in amplifier, as illustrated in Fig. 3a. To avoid the influence of $t_{FM}$ on orbital absorption and self-torque, we fix $t_{FM}$ and directly measure SOT effective field instead of $V_S/V_A$, where $V_S$ and $V_A$ are symmetric and antisymmetric component. The signal $V_{DC}$ induced by y-polarized OAM and Oersted field in y-direction can be expressed by [35,36]

$$V_{DC} = \frac{I_{RF}\Delta R \sin 2\varphi \cos \varphi}{2\alpha(2H_{res} + M_{eff})} \left( S(H)H_{DL} + \sqrt{1 + \frac{M_{eff}}{H_{res}}} A(H)H_{FL+Oe} \right) \quad (2)$$

where the symmetric Lorentzian $S(H) = W^2/[(H - H_{res})^2 + W^2]$ arises from damping-like torque $H_{DL}$, and antisymmetric Lorentzian $A(H) = W(H - H_{res})/[(H - H_{res})^2 + W^2]$ is induced by field-like torque and Oersted field $H_{FL+Oe}$, $H_{res}$ is resonance field and $W$ is the half-width-at-half-maximum linewidth. $M_{eff}$ is effective magnetization fitted by the Kittel formula $2\pi f = \gamma\sqrt{H_{res}(H_{res} + M_{eff})}$, and $\gamma$ is gyromagnetic ratio. $\alpha$ is Gilbert damping coefficient fitted by $W = 2\pi f \alpha/\gamma + W_0$. $\Delta R$ is in-plane ADMR arising from AMR and other mechanisms obtained by $R(\varphi) = R_0 + \Delta R \cos^2 \varphi$. $I_{RF}$ is the amplitude of RF current measured by comparing the resistance changes due to the Joule heating of RF and DC currents. More details are shown in Supplemental Material S1. Fig. 3b shows typical $H$ dependence of $V_{DC}$, the fitted $S(H)$ contribution $V_S$ and $A(H)$ contribution $V_A$ of sample Ru(20)/Co(5) at $f$=14 GHz and $\varphi$=35°. The significant $V_S$ signal indicates the existence of $H_{DL}$. To exclude the influence of antisymmetric current on the signal, $\varphi$ dependence of $V_S$ is extracted (see Fig. 3c). The fitting curve is mainly contributed by $\sin 2\varphi \cos \varphi$, while $\sin 2\varphi$ contribution caused by z-polarized Oersted field is much less. Considering the weak SHE of Ru [37], $H_{DL}$ can be regarded as the sign of orbital torque. To normalize orbital torque, we calculated SOT efficiency by

$$\xi_{DL}^{E} = \frac{2e}{\hbar} \frac{\mu_0 M_s t_{FM} H_{DL}}{E} \quad (3)$$

where $E$ is the electric field obtained by $I_{RF}R_0$, and $M_s$ is the saturation magnetization. Fig. 3d reveals the $\xi_{DL}^{E}$ of Ru(t)/Co(5) with various $t_{Ru}$, and significant orbital torque is observed. The $\xi_{DL}^{E}$ increases from 0.5×10$^5$ Ω$^{-1}$m$^{-1}$ to about 5×10$^5$ Ω$^{-1}$m$^{-1}$ and gradually saturates, which can be described by drift-diffusion model

$$\xi_{DL}^{E} = \sigma_{OH}^{eff}\left(1 - \text{sech}\frac{t_{NM}}{\lambda_{NM}}\right) + \xi_{DL,0}^{E} \quad (4)$$

where $\lambda_{NM}$ is the diffusion length of orbital current in NM, $\sigma_{OH}^{eff}$ is effective orbital Hall conductivity, and $\xi_{DL,0}^{E}$ is self-torque or interfacial Rashba contribution. From Fig. 3d, the drift-diffusion model well fits various $\xi_{DL}^{E}$, where fitted $\lambda_{Ru}$=13.1±2.4 nm and $\sigma_{OH}^{eff}$=5.45×10$^5$ Ω$^{-1}$m$^{-1}$, which is significantly higher than that of Pt/Co [38]. And the $\xi_{DL,0}^{E}$ of Ru(t)/Co(5) is negligible. Importantly, the strong orbital torque indicates the existence of large orbital current generated by Ru.

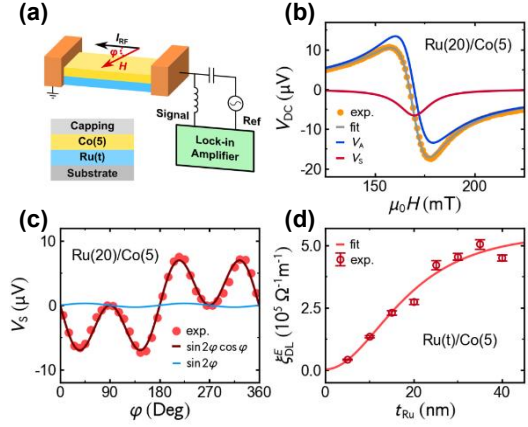

Fig. 3. (a) Schematic of the ST-FMR measurement setup. (b) Magnetic field dependence of the rectified voltage $V_{DC}$ for Ru(20)/Co(5) at $f$=14 GHz and $\varphi$=35 °, decomposed into symmetric ($V_S$, red) and antisymmetric ($V_A$, blue) Lorentzian components. (c) Angular dependence of $V_S$ fitted by a function with $\sin 2\varphi \cos \varphi$ contribution (dark red line) and $\sin 2\varphi$ contribution (light blue line). (d) Damping-like torque efficiency $\xi_{DL}^E$ as a function of Ru thickness $t_{Ru}$ for Ru(t)/Co(5) series, and fitted by the drift-diffusion model (solid line).

The contradiction between the strong orbital torque and the vanishing OMR validates the failure of SMR-like theory for OAM. To analyze the discrepancy between the predicted OMR based on SMR theory and experimental results, we construct an OMR model referencing SMR model [24], which can be expressed by

$$\frac{\Delta R}{R_0} = \theta_{OH}^2 \frac{\lambda_{NM}}{t_{NM}} \frac{\tanh^2(t_{NM}/2\lambda_{NM})}{1 + t_{FM}\rho_{NM}/t_{NM}\rho_{FM}} \left[ \frac{g_R}{1 + g_R \coth(t_{NM}/\lambda_{NM})} - \frac{g_F}{1 + g_F \coth(t_{NM}/\lambda_{NM})} \right] \quad (5)$$

where orbital Hall angle $\theta_{OH}$ and $\lambda_{NM}$ can be obtained by orbital torque measurement. We initially assume an ideal Ru/Co interface as interfacial orbital transparency $T_{orb} \approx 1$. Thus, $\theta_{OH}$ can be approximated by effective orbital Hall angle $\theta_{OH}^{eff} = \rho_{NM}\sigma_{OH}^{eff} = T_{orb}\theta_{OH}$. And in the square brackets of the formula, the first and second terms involving $g_R$ and $g_F$ denote the absorption of orbital current when $\boldsymbol{\sigma}_{orb} \perp \boldsymbol{M}$ and $\boldsymbol{\sigma}_{orb} \parallel \boldsymbol{M}$, respectively. Where $g_R = 2\rho_{NM}\lambda_{NM}\text{Re}[G_{mix}^{orb}]$ and $g_R \gg 1$ for ideal interface, $g_F = \eta\rho_{NM}\lambda_{NM}/\rho_{FM}\lambda_{FM} \coth(t_{FM}/\lambda_{FM})$ and $g_F$=0 for ideal situation, $G_{mix}^{orb}$ means orbital mixing conductance and $\eta$ is orbital absorption efficiency. Here, we assume $g_R$=10 and $g_F$=0.1 by analogy with SMR. The predicted $\Delta R/R_0$ by the SMR-like model are plotted in Fig. 2c. Clearly, the predicted OMR significantly deviates from the experiment data. This result indicates that the OMR based on SMR theory should be detected and the absence of observable OMR stems from the weak dependence of orbital reflection on $\boldsymbol{M}$.

The limitation of the SMR-like OMR model stems from the underestimation of $g_F$. For SMR, the strong coupling between spin and magnetic moment results in strong interfacial spin reflection and weak spin absorption when $\boldsymbol{\sigma}_{spin} \parallel \boldsymbol{M}$, leading to a small value of $g_F$. In contrast, for OMR, the intense orbital absorption by FM when $\boldsymbol{\sigma}_{orb} \parallel \boldsymbol{M}$ implies that $g_F$ approaches the magnitude of $g_R$. To illustrate this, Fig. 4a plots the simulated impact of $g_F$ on OMR, where $\theta_{OH}$=0.1, $g_R$=1, $\lambda_{NM}$=10 nm, $t_{FM}$=5 nm, $\rho_{NM}$=20 μΩ·cm and $\rho_{FM}$=50 μΩ·cm. Clearly, OMR reduces significantly with $g_F$ increasing. Furthermore, considering the interference of intrinsic MR from FM layer and current shunting on $\Delta R/R_0$, the $t_{NM}$-dependence of OMR becomes submerged within these contributions and difficult to separate, as shown in Fig. 4b.

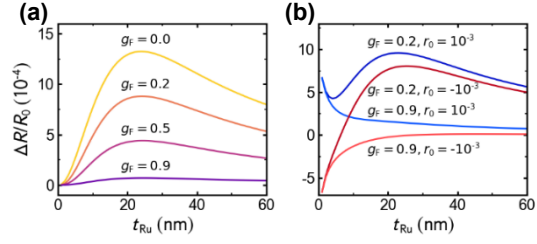

Fig. 4. (a) Calculated OMR as a function of $t_{NM}$ for varying orbital absorption $g_F$, assuming $\theta_{OH}=0.1$, $g_R=1$, $\lambda_{NM}=10$ nm, $t_{FM}=5$ nm, $\rho_{NM}=20$ μΩ·cm and $\rho_{FM}=50$ μΩ·cm. (b) Simulation of the total MR ratio combining OMR and intrinsic FM magnetoresistance contribution $r_0$. A large $g_F$ renders the OMR signal indistinguishable from background contributions from FM.

To further investigate the orbital torque and OMR in NM/FM bilayers, we also examine Ni-based bilayers commonly used in orbital-related studies, including Ru(t)/Ni(8), Ni(8)/Ru(t), and Ni(8)/Ti(t). In these samples, we consider both the deposition order and the choice of NM, where both Ti and Ru possess strong OHE but negligible SHE [37,39]. Importantly, the ADMR analysis for Ni-based films differs from that used previously, because of the significant deviation of the ADMR curves from $\cos^2 \beta$ dependence (see Supplemental Material S2). This behavior originates from the growth-induced texture of Ni, where crystalline anisotropy causes higher-order contributions in Ni-based films [40]. As a result, compared with the above approach using $\Delta R = R_\perp - R_\parallel$, the function $R = R_0 + \sum_n \Delta R_{2n} \cos^{2n} \beta$ including higher-order terms is necessary. Here, $\Delta R_2$ term represents the underlying OMR and SMR or other MR mechanism, high-order terms $\Delta R_4$, $\Delta R_6$ and $\Delta R_8$ are induced by crystal anisotropy (more details see Supplemental Material S2). To demonstrate the necessity of these higher-order terms, we compare the MR ratios obtained from two methods in Pt(t)/Ni(8) series. Only the $\Delta R/R_0$ extracted from the high-order fitting, that is $\Delta R_2/R_0$, follows the expected $t_{NM}$-dependence of the SMR effect [24] (see Supplemental Material S3).

Subsequently, the high-order analysis is used for Ru(t)/Ni(8), Ni(8)/Ru(t), and Ni(8)/Ti(t) series to extract $t_{NM}$-dependence $\Delta R/R_0$ as shown in Fig. 5a-c. For Ni(8)/Ru(t) and Ni(8)/Ti(t) with Ni grown at the bottom, the $\Delta R/R_0$ are negative and monotonically decay from $4\times10^{-4}$ to $5\times10^{-5}$ with $t_{NM}$ similar to Ru(t)/Co(5) (see Fig. 5b, c). These behaviors are well reproduced by the shunting model (solid lines in Fig. 5b, c), indicating that no OMR is observed. In contrast, as shown in Fig. 5a, the $\Delta R/R_0$ for Ru(t)/Ni(8) exhibits a sign reversal from negative to positive at $t_{Ru}=10$, and it increased first and then decreased, forming a peak at $t_{Ru}=15\sim20$. Although Ru(t)/Ni(8) has a peak-like MR, this peak arises from a sign reversal of the $\Delta R/R_0$ induced by the texture of NM layer [32] rather than OMR. The lower-order MR term $\Delta R_2$ of Ni film is sensitive to the structural details including buffer layer, which might complicate the analysis of $\Delta R/R_0$. One evidence is the influence of buffer layer on $\Delta R/R_0$ in Ru(t)/Ni(8) and Pt(t)/Ni(8), in which the samples with and without amorphous $SiO_2$ buffer layer have different $\Delta R/R_0$ (see Supplemental Material S4). For the Ni(8)/Ru(t) samples, since Ni is deposited first, its structural properties remain constant by the thickness of Ru overlayer, and $\Delta R/R_0$ therefore shows a stable and monotonic decay dominated by current shunting. However, in the Ru(t)/Ni(8) samples, the strong structural sensitivity of Ni means that variations in the underlying Ru layer influence the Ni texture, resulting in MR sign reversal. Moreover, the MR of Ni(8)/Ti(t), Ni(8)/Ru(t), and Ru(5)/Ni(8) are all negative, which further supports that the sign reversal in Ru(t)/Ni(8) originates from structural evolution. In summary, no trace of OMR is observed in these series.

However, significant orbital torques are observed in all three series (see Fig. 5e–g). For Ru(t)Ni(8), The $\xi_{DL}^E$ increases and gradually saturates with $t_{Ru}$. For Ni(8)/Ru(t) and Ni(8)/Ti(t), the $\xi_{DL}^E$ decreases with $t_{Ru}$ or $t_{Ti}$ due to the reversed deposition order, and Ni(8)/Ti(t) does not have saturation behavior because of longer $\lambda_{Ti}$. Furthermore, all $t_{NM}$-dependent $\xi_{DL}^E$ are well described by the drift-diffusion model, the extracted parameters are summarized in Tab. 1. The fitted $\lambda_{Ru}$ of Ru(t)/Ni(8) and Ni(8)/Ru(8) are consistent with that obtained in the Ru(t)/Co(5) series, and the $\lambda_{Ti}$ determined to be 44.1±2.3 nm is consistent with previously reported large orbital diffusion length [19]. Moreover, all fitted curves of Ni-based series deviate from 0 at $t_{NM}$=0, meaning large $\xi_{DL,0}^E$, which might arise from larger self-torque or interfacial Rashba effect in Ni-based bilayers. Notably, Ru(t)/Ni(8) exhibits a higher SOT efficiency than Ni(8)/Ru(t), which may originate from the difference in interfacial orbital transparency caused by the deposition order, as previous research reported [19,41]. In addition, Ru(t)/Co(5) owns higher $\xi_{DL}^E$ compared with Ru(t)/Ni(8), which might originate from a higher orbital transparency at the Ru/Co interface, and strong orbital torques in Co have also been reported previously [42].

Based on orbital torque measurement, we also predict their OMR based on SMR theory. Due to the weaker $\xi_{DL}^E$ in Ru(t)/Ni(8) and Ni(8)/Ru(t) compared with Ru(t)/Co(5), the assumption of ideal interface and $g_R \gg 1$ do not hold. Thus, assume $\theta_{OH}$=0.12 based on Ru(t)/Co(5), and $g_R$ can be calculated by $\theta_{OH}^{eff} = T_{orb}\theta_{OH}$ and $T_{orb} = g_R/(1 + g_R)$ [43]. For Ni(8)/Ti(t), we still assume ideal interface, use $\theta_{OH,eff}$ as $\theta_{OH}$ and set $g_R$=10 like in Ru(t)/Co(5). As the dashed lines plotted in Fig. 5a-c the OMR predicted by the model is remarkably large and apparently deviates from experiment. This further supports our hypothesis that the signal of OMR is extremely weak.

To further demonstrate that the SOT comes from orbital torque and MR is independent of the interaction between orbital and magnetic moment, we measure Ru(t)/Py(5) series using Py with weak orbital torque [19,28] as FM. For ADMR, $\Delta R/R_0$ is negative and monotonically decays from 1×10$^{-3}$ to 5×10$^{-5}$ with $t_{NM}$, which is similar to previous results and might arise from geometric size effects and other intrinsic mechanisms [31,33]. Moreover, the evolution of $\Delta R/R_0$ with $t_{Ru}$ can also be well described by the shunting model, as the solid line shown in Fig. 5d. From Fig. 5h, the $\xi_{DL}^E$ of Ru(t)/Py(5) is small, which indicates orbital torque dominated SOT in the previous experiment. These results indicate that the appearance and variation of MR are not correlated with the orbital torque. Instead, the MR mainly arises from interface-related effects, geometric size effect, and other mechanisms inherent to the FM layer.

Tab. 1. The fitted results and parameters of drift-diffusion model of orbital torque, shunting model of MR and SMR-like OMR model.

| unit | $\lambda_{NM}$ nm | $\sigma_{OH}^{eff}$ 10$^5$ Ω$^{-1}$m$^{-1}$ | $\rho_{NM}$ μΩ·cm | $\rho_{FM}$ μΩ·cm | $r_0$ 10$^{-4}$ |
|---|---|---|---|---|---|
| Ru(t)/Co(5) | 13.1±2.4 | 5.45±0.48 | 22.4 | 164 | 66.8±3.3 |
| Ru(t)/Ni(8) | 14.8±2.3 | 1.77±0.22 | 22.4 | 37.2 | - |
| Ni(8)/Ru(t) | 11.3±2.0 | -0.87±0.08 | 20.9 | 64.5 | -10.0±0.64 |
| Ni(8)/Ti(t) | 44.1±2.3 | -1.14±0.07 | 71.3 | 64.5 | -6.04±0.25 |
| Ru(t)/Py(5) | - | - | 22.4 | 229 | -104±5.8 |

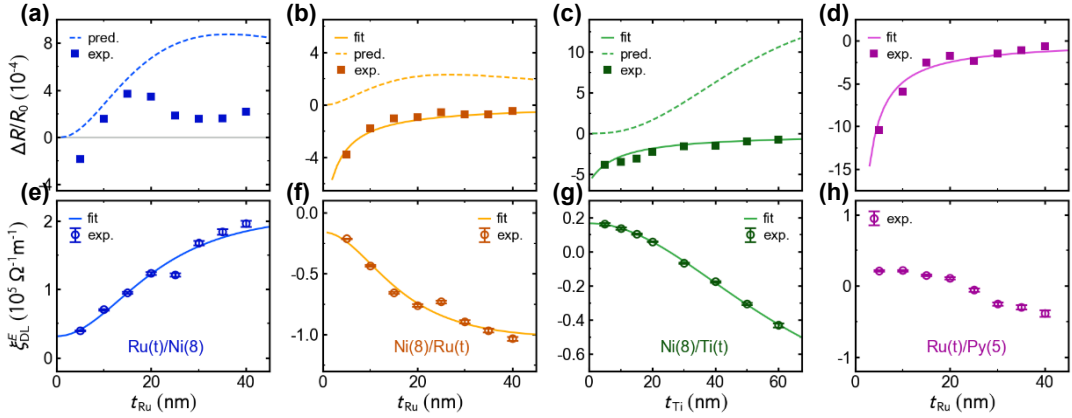

Fig. 5. (a-d) The $t_{NM}$-dependence of MR ratios $\Delta R/R_0$ for (a) Ru(t)/Ni(8) (b) Ni(8)/Ru(t), (c) Ni(8)/Ti(t) and (d) Ru(t)/Py(5) series. The solid lines represent the fitting of shunting model, consistently reproducing the data without invoking OMR contributions. And the dashed lines indicate the predicted OMR based on SMR theory. (e-h) Corresponding orbital torque efficiency $\xi_{DL}^{E}$ for the respective series. Solid lines denote fits to the drift-diffusion model.

Based on the above results and discussion, we emphasize that Ni-based bilayers are not ideal candidates for investigating orbital or even spin-related effects. For instance, we observed that the texture of Ni films introduces MR contributions and generates SMR-like artifacts driven by texture evolution, which are sensitive to the buffer layer thickness and structure. Furthermore, in SOT measurements, Ni-based bilayers exhibit significant self-torques or interfacial Rashba effects. These contributions can reduce or enhance orbital torque, which may interfere with the evaluation of OHE, particularly in experiments lacking $t_{NM}$-dependence or where $t_{NM}$ is below $\lambda_{NM}$. A typical controversy highlights this issue: whether the positive SOT observed in Ta/Ni originates from the orbital torque [17,44] or self-torque [45].

In conclusion, our measurements show a contrast between giant orbital torques and the absence of observable OMR in Ru/FM and Ti/FM bilayers. The variation of MR is entirely dominated by the intrinsic MR of FM layer combined with current shunting, with negligible contribution from orbital transport. We attribute this to the relaxation mechanisms of orbital current deviating from spin: orbital currents are strongly absorbed by FM regardless of whether the orbital polarization is parallel or perpendicular to magnetization. This isotropic bulk absorption suppresses the reflection anisotropy essential for SMR-like effects. Our findings challenge the direct spin-orbital analogy for orbital phenomena and suggest that re-examining the mechanism of orbital-related effects.

**Acknowledgments**

**This work was supported by the CAS project for Yong Scientists in Basic Research (Grant No. YSBR-030), the National Natural Science Foundation of China (Grant No. 12474102).**